\documentclass[twocolumn,aps,prd,nofootinbib]{revtex4-1}
\usepackage[colorlinks=true,linkcolor=black,citecolor=blue,urlcolor=black]{hyperref}
\usepackage{graphicx,soul}
\usepackage{subfigure}
\usepackage{natbib}
\usepackage{epsfig}
\usepackage{amstext,amsmath,amsfonts,amssymb,amsthm}
\usepackage[format=plain,justification=centerlast]{caption}
\usepackage{color}

\usepackage{verbatim}
\usepackage{lipsum}
\usepackage{color,array,wrapfig}
\usepackage{caption}
\usepackage{notes2bib}
\begin{document}

\title{Enhanced diffusion, swelling and slow reconfiguration of a single chain in non-Gaussian active bath}
\author{Subhasish Chaki and Rajarshi Chakrabarti*}
\affiliation{Department of Chemistry, Indian Institute of Technology Bombay, Mumbai, Powai 400076, E-mail: rajarshi@chem.iitb.ac.in}
\date{\today}

\begin{abstract}
\noindent A prime example of non-equilibrium or active environment is a biological cell. In order to understand in-vivo functioning of biomolecules such as proteins, chromatins, a description beyond equilibrium is absolutely necessary. In this context, biomolecules have been modeled as Rouse chains in Gaussian active bath. However, these non-equilibrium fluctuations in biological cells are non-Gaussian. This motivates us to take a Rouse chain subjected to a series of pulses of force with finite duration, mimicking run and tumble motion of a class of micro-organisms. Thus by construction, this active force is non-Gaussian. Our analytical calculations show that the mean square displacement (MSD) of center of mass (COM) grows faster and even shows superdiffusive behavior at higher activity, supporting recent experimental observation on active enzymes (A.-Y. Jee, Y.-K. Cho, S. Granick, and T. Tlusty, Proc. Natl. Acad. Sci. 115, 10812 (2018)), but chain reconfiguration is slower. The reconfiguration time of a chain with $N$ monomers scales as $~N^\sigma$, where the exponent $\sigma \approx 2$. In addition, the chain swells. We compare this activity-induced swelling with that of a Rouse chain in a Gaussian active bath. In principle, our predictions can be verified by future single molecule experiments. 
\end{abstract}

\maketitle

\begin{figure}[b]
\centering
 \includegraphics[width=0.35\textwidth]{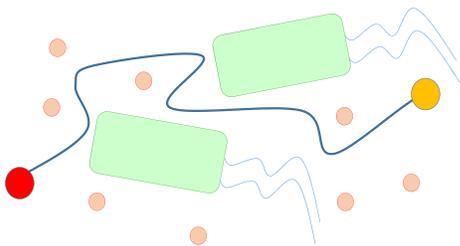}
 \caption{Schematic of the model (not to scale): Single Rouse chain (blue) is immersed in a bath of bacteria (green). The orange balls represent the fluid molecules. Yellow and red ball represent the dyes attached at the two ends of the chain.}
 \label{fig:pic1}
 \end{figure}

\noindent Attempts to understand the dynamics of long chain biomolecules have gained extensive research interests in recent years. Inside a living cell, they are not solely driven by thermal fluctuations but rather fueled by the energy released upon hydrolysis of ATP into directed motion. This continuous conversion of chemical energy breaks the microscopic detailed balance and drives the system out of equilibrium (even in the absence of external forcing) \cite{gladrow2016broken}.  Examples of such systems are molecular motors \cite{sonn2017scale},  active membranes \cite{park2010measurement}, motile bacteria \cite{wu2000particle} or self-propelled Janus particles \cite{gomez2016dynamics}. Very recently people have also started viewing enzymes as active matter \cite{jee2018enzyme,jee2018catalytic}. Inspired by these active processes, a series of simulation studies have also been performed to build models for active polymers where monomers are treated as active Brownian particles (ABP) \cite{chelakkot2014flagellar,anand2018structure,bianco2018globule,sarkar2016coarse}. However, the dynamics of passive systems in active environment exhibit even more fascinating features \cite{maggi2014generalized,chaki2018,nandi2017nonequilibrium}. A representative example of such system is an eukaryotic cell where proteins are constantly driven out of equilibrium by a range of active processes in addition to thermal fluctuations from surroundings \cite{hu2016dynamics,metzler2016protein}. This motivated people to come up with polymer based models in active bath that showed enhanced diffusion of tagged monomers and subsequent swelling of the chain \cite{kaiser2014unusual,samanta2016chain,osmanovic2017dynamics,vandebroek2015dynamics, eisenstecken2017internal,shin2015facilitation,ghosh2014dynamics, osmanovic2018dynamics}.  In all of these theoretical studies, the active noise was modeled as Gaussian random variable.  However, this Gaussian approximation works well when the density of the active particles is very low \cite{szamel2014self,zakine2017stochastic} or the local relaxation time $(\tau_{\text{relax}})$ is greater than the correlation time $(\tau_A)$ of the active noise \cite{argun2016non}. For harmonic motion,  $\tau_{\text{relax}}=\frac{\gamma}{k}$ where $\gamma$ is friction coefficient and $k$ is spring constant. But if the density of active particles is high, resulting comparable or even higher $\tau_A$ than $\tau_{\text{relax}}$, then the active forces are no longer Gaussian. For example, Krishnamurthy $et\,\,al.$ have experimentally  shown that the displacement statistics of a colloidal particle in a time-varying optical potential across bacterial baths becomes increasingly non-Gaussian with the activity \cite{krishnamurthy2016micrometre}, Toyota $et\,\,al.$ have experimentally shown that a tracer bead immersed in  acto-myosin network is subjected to active non-Gaussian fluctuations \cite{toyota2011non}. However, theoretical study on biomolecular dynamics in a bath of bacteria and molecular motors is lacking. Such studies are extremely important as protein folding might be facilitated by the presence of active fluctuations \cite{harder2014activity} and non-Gaussian fluctuations could play a pivotal role in it. 

\noindent To monitor non-equilibrium folding dynamics of a peptide or a protein, people rely on single-molecule nanosecond fluorescence correlation spectroscopy (nsFCS) that employs F$\ddot{o}$rster resonance energy transfer (FRET) between a pair of dyes attached at different locations along the backbone \cite{soranno2012quantifying}. Fluctuations in the distance between  FRET pairs is related to the fluctuations in fluorescence intensities and the characteristic timescale for these fluctuations referred to reconfiguration time. A good estimate of that is determined by fitting the long time decay of intensity autocorrelation function \cite{makarov2010spatiotemporal}. This reconfiguration time provides a general notion of the rate of change of the conformations of a polymer.  Investigations along this direction, are highly important as reconfiguration time is an essential measure of characteristic timescale of conformal changes of a single chain driven by active fluctuations. 

\noindent In order to understand dynamics of biomolecules in a bath of active particles, we consider a Rouse polymer subjected to a thermal and a non-Gaussian active force arising from this bath of active particles (such as bacteria and active enzymes). A very good description of this active force is shot noise that explicitly accounts for the run-and-tumble motion of active particles. In case of such shot noise driven processes, absence of sufficient number of random kicks at every instance of time leads to the violation of central limit theorem and hence results non-Gaussian statistics. Our model shows that collisions from active particles enhance mean square displacements (MSDs) of the chain. In addition, it predicts slower reconfiguration in the presence of active noise. Like an equilibrium Rouse chain, it scales as $N^2$ where $N$ is the number of monomers. We also analyze the swelling behavior of a Rouse chain in the presence of non-Gaussian active noise and compare this with a Rouse chain in a Gaussian active bath \cite{samanta2016chain}. 

\noindent A coarse grained description of a biomolecule is a  linear chain with $N$ beads (monomers) connected by harmonic springs with spring constant $k\left(=\frac{3k_B T}{b^2}\right)$, where the Kuhn length, $b$, gives the length scale of flexibility of the polymer. The origin of this spring force is purely entropic in nature. Presence of active particles enters in the Rouse model through an extra noise, ${\eta_A}(n,t)$ in addition to thermal noise, ${\xi_T}(n,t)$ arising from solvent motion. Here ${\xi_T}(n,t)$ and ${\eta_A}(n,t)$ are uncorrelated as they associated with completely different time scales. 
\noindent The equation of motion for the $n^{th}$ monomer in the continuum limit is given by
\begin{equation}
\gamma\frac{\partial{R_n}(t)}{\partial{t}}=k\frac{\partial^2{R_{n}(t)}}{\partial{n^2}}+{\xi_T}(n,t)+{\eta_A}(n,t)
\label{eq:rouse-model}
\end{equation}
\noindent Where $R_n(t)$ is the position of $n^{th}$ monomer and $\gamma$ is the friction coefficient of the solvent. The thermal force, ${\xi}(n,t)$ is assumed to be stationary, Markovian, and Gaussian with zero mean and variance,
\begin{equation}
   \left<\xi_{T\alpha}(n,t^{\prime})\xi_{T\beta}(m,t^{\prime\prime})\right>=2 \gamma k_B T \delta_{\alpha\beta}\delta(n-m)\delta(t^{\prime}-t^{\prime\prime})
\label{eq:random-forcerouse}
\end{equation}
\noindent We describe ${\eta_A}(n,t)$ later.  Decomposing $R_n(t)$ into normal modes ($X_p(t)$) as ${R_n}(t)={X_0} + 2 \sum\limits_{p=1}^\infty U_{pn} X_p(t)$, one can show $X_p(t)$ has following dynamics,
\begin{equation}
\gamma_p \frac{d{X_p}(t)}{d{t}}=-k_p X_{p}(t)+\xi_{T,p}(t)+\eta_{A,p}(t)
\label{eq:rouse-mode}
\end{equation}
\noindent Since there is no external force applied on the chain, $U_{pn}=cos(\frac{p \pi n}{N})$. The relaxation time for the $p^{th}$ normal mode is $\tau_p=\frac{\gamma N^2 b^2}{3 k_BT\pi^2 p^2}$. The thermal noise immediately follows,
\begin{equation}
\left<\xi_{T,p\alpha}(t^{\prime})\xi_{T,q\beta}(t^{\prime\prime})\right>=2 \gamma_p k_B T \delta_{\alpha\beta}\delta_{pq}\delta(t^{\prime}-t^{\prime\prime})
\label{eq:randomforce_modes}
\end{equation}
\noindent The strength of the thermal noise satisfies fluctuation-dissipation theorem (FDT) that maintains an equilibrium between the system and the thermal bath. Here we like to point out that each normal mode is Markovian and its correlation is a single exponential. Thus, ${R_n}(t)$ is a linear superposition of Markov processes and thus non-Markovian in nature. As a matter of fact, even if one does not explicitly consider viscoelastic  polymer described by a generalized Langevin model, a simple Rouse description carries non-Markovianity \cite{dua2011non}. 
\noindent The equation (\ref{eq:rouse-mode}) is structurally same as an over-damped Brownian particle trapped in a harmonic well in the presence of active noise,
\begin{equation}
 \gamma\frac{dx}{d{t}}=-k x(t)+\xi_T(t)+\eta_A(t)
\label{eq:langevin_active}
\end{equation}

\begin{figure}[h]
\begin{center}
 \includegraphics[width=0.45\textwidth]{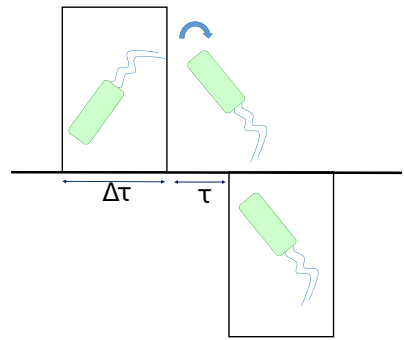}
 \caption{Schematic diagram of series of active pulses generated by bacteria (green).}
 \label{fig:pic}
\end{center}
 \end{figure}

\begin{figure*}[ht]
\begin{center}
\begin{tabular}{cc}
\resizebox{9.25cm}{!} {\includegraphics*[width=9.25cm]{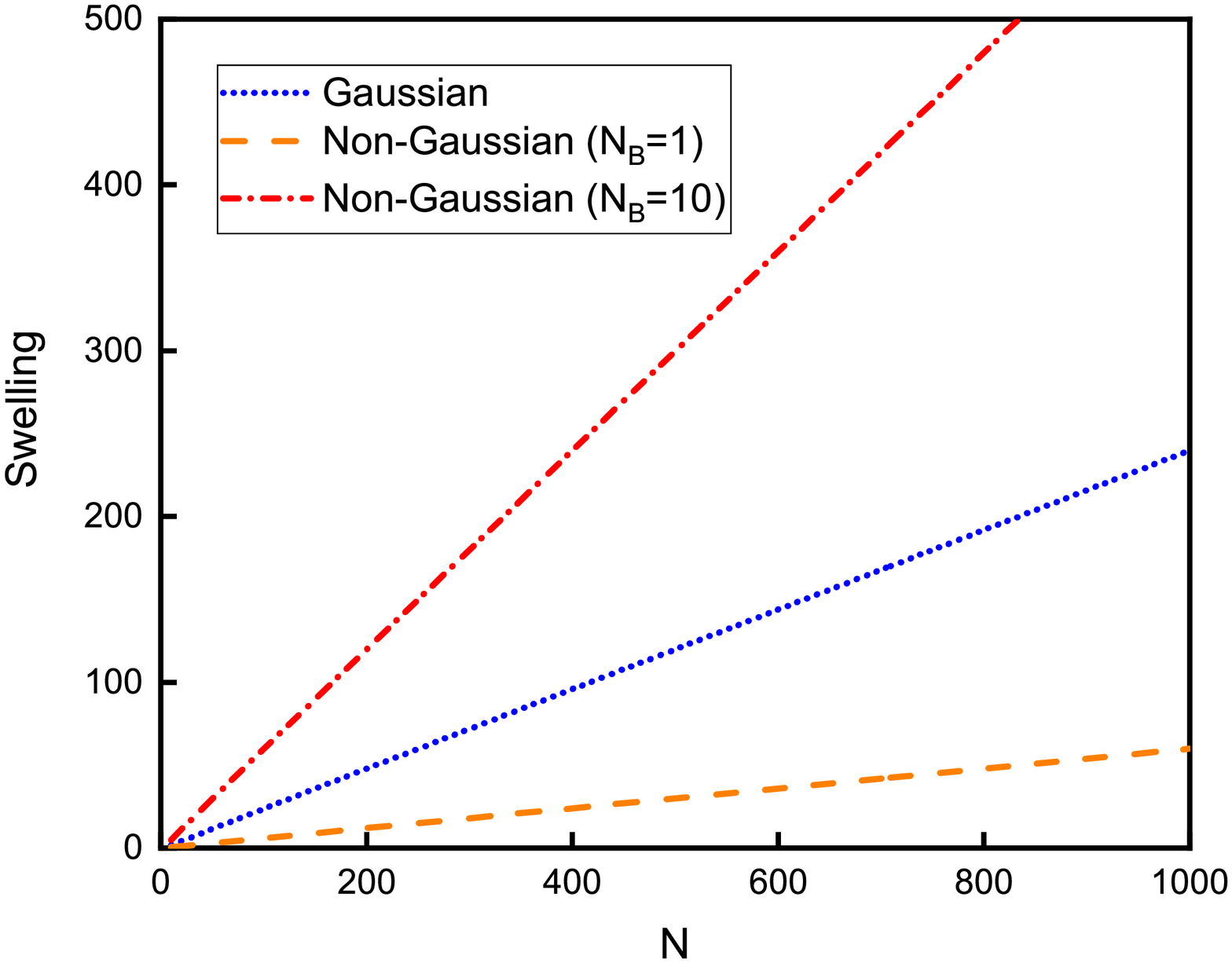}} &
\resizebox{9.25cm}{!} {\includegraphics*[width=9.25cm]{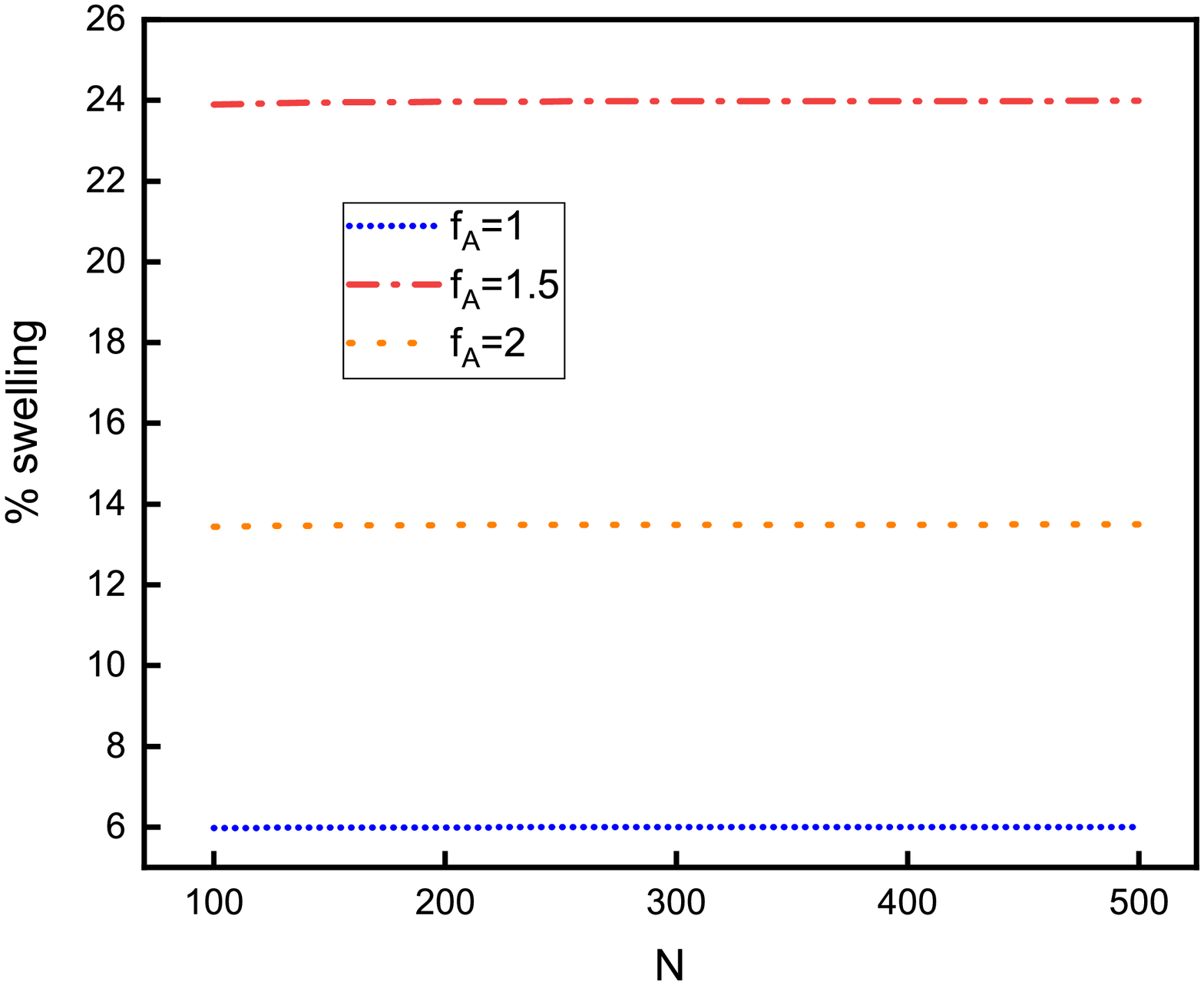}}\\
(a) & (b)  \\
\end{tabular}
\end{center}
\caption{(a) Plot of swelling versus $N$ for Rouse chain. The values of  parameters used are   $b=1, \gamma=1, k_B=1, T=1, \Delta\tau=1, \tau=1, f_A=1$  (for non-Gaussian chain) and $b=1, \gamma=1, k_B=1, T=1, \tau_A=1, C=1$ (for Gaussian chain) respectively. (b)  Plot of \%swelling versus $N$ for Rouse chain. The values of the parameters used are   $b=1, \gamma=1, k_B=1, T=1, \Delta\tau=1, \tau=1, N_B=1 $}
\label{fig:swelling}
\end{figure*}

\begin{figure*}[ht]
\begin{center}
\begin{tabular}{cc}
\resizebox{9.25cm}{!} {\includegraphics*[width=9.25cm]{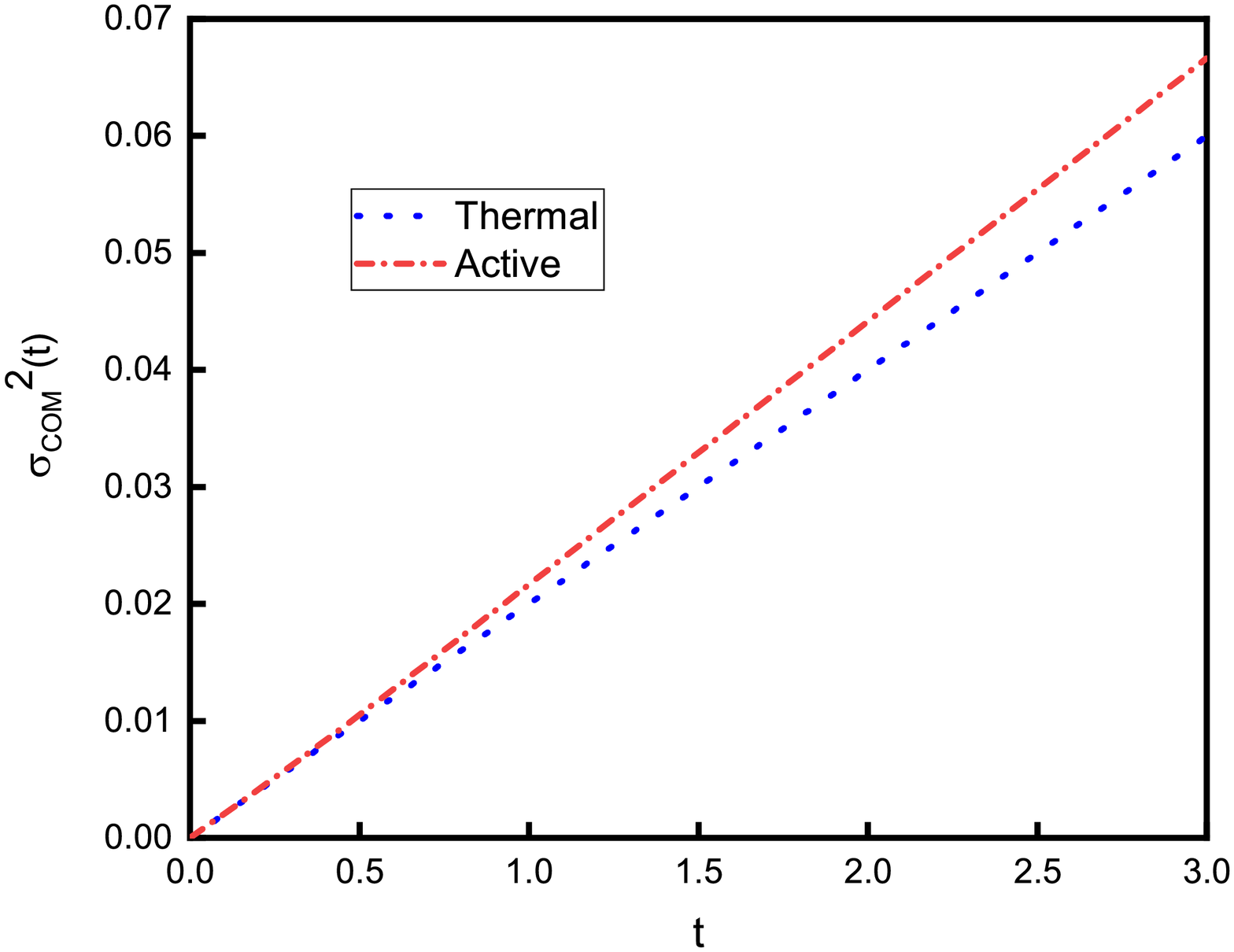}} &
\resizebox{9.25cm}{!} {\includegraphics*[width=9.25cm]{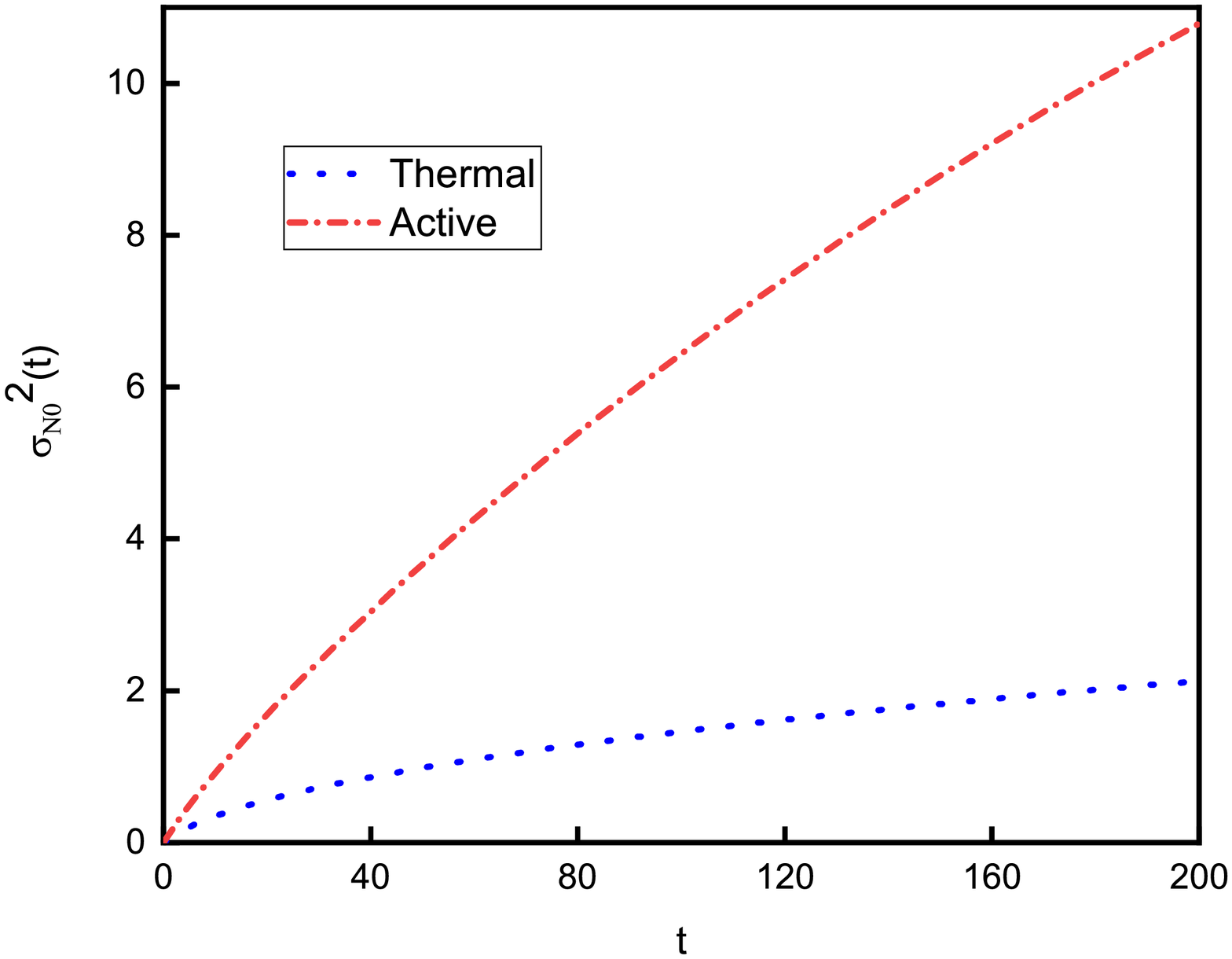}}\\
(a) & (b)  \\
\end{tabular}
\end{center}
\caption{Plot of MSD versus $t$ for a Rouse chain of length $N=100$: (a) for the end-to-end distance. (b) for the COM. The values of the parameters for COM used are  $b=1, \gamma=1, k_B=1, T=1, \Delta\tau=1, \tau_A=1, N_B=1, f_A=1$ and for end-to-end distance  $b=1, \gamma=1, k_B=1, T=1, \Delta\tau=1, \tau_A=1, N_B=1, f_A=0.02$}
\label{fig:msd}
\end{figure*}

\noindent To comment the nature of $\eta_A(t)$, we go back to  run-and-tumble strategy of bacteria motion, when a bacterium swim straight over a given time (run) and exert a force, $f(t)$ of duration $\Delta \tau$ on polymer until it change the direction (tumble)  \cite{pohl2017inferring}. Recent experiments  \cite{jee2018enzyme,jee2018catalytic} suggest that enzyme migration towards substrate involves a similar run-and-tumble motion. Thus $f(t)=f_A[\Theta(t)-\Theta(t-\Delta \tau)]$ where $f_A$ is the magnitude of the force $f(t)$ and $\Theta(t)$ is an unit step function. Consequently, the active noise $\eta_A(t)$ is represented by a series of random pulses and modeled as shot noise, $\eta_A(t)=\sum\limits_i h_i f(t-t_i)$, where $h_i = \pm 1$ is the sign of the $i^{th}$ pulse accounts for the direction of the force and $t_i$ is the time at which the $i^{th}$ pulse started. Here, $t_i$ is a exponentially distributed random variable with a constant rate $\nu=\frac{N_B}{\tau + \Delta \tau}$. $N_B$ is the number of bacteria and there is no overlapping between the pulses produced by different bacteria. $\tau$ is the average waiting time between the end of one pulse to the next time it starts. The direction of the force on polymer should be equal and independent from each other. Thus $h_i$ is symmetric  with unit strength and correlation $ \left<h_i h_j\right>=\delta_{ij}$. For our active Rouse chain, $\left<\eta_{A,p\alpha}(t)\eta_{A,q\beta}(0)\right>=\frac{\nu f_A^2N\delta_{\alpha\beta}\delta_{pq}  (\Delta \tau-|t|)\Theta(\Delta \tau-|t|)}{2}$ \cite{supplemental}. However, for active Ornstein-Uhlenbeck process (AOUP), we have to consider $\eta_A(t)$ as Gaussian random variable with zero mean and exponentially decaying correlation $\left\langle \eta_A(t) \eta_A(t^{\prime}) \right\rangle=Ce^{-\frac{|t-t^{\prime}|}{\tau_A}}$ in Eq.(\ref{eq:langevin_active}) \cite{samanta2016chain}. If we consider the time difference $|t-t^{\prime}|$ within $\tau_{\text{relax}}$ and $\tau_A >\tau_{\text{relax}}$, we can approximate $e^{-\frac{|t^{\prime}-t^{\prime\prime}|}{\tau_A}} \approx \left(1-\frac{|t^{\prime}-t^{\prime\prime}|}{\tau_A}\right)$ which is structurally same as the noise autocorrelation for run and tumble motion, $(\Delta \tau-|t^{\prime}-t^{\prime\prime}|)\Theta(\Delta \tau-|t^{\prime}-t^{\prime\prime}|)$. Thus for large correlation time $(\tau_A)$ introduced by the active noise, a transition from Gaussian to non-Gaussian active noise is expected as predicted by Volpe $et\,\,al.$ \cite{argun2016non}.
\noindent Using Campbell's theorem \cite{balakrishnan2008elements}, \noindent time correlation function of normal modes of the Rouse chain can be written as \cite{supplemental},
\begin{widetext}
\begin{equation}
\begin{split}
\left<X_{p\alpha}(t)X_{q\beta}(0)\right>&=\frac{k_B T}{k_p}\delta_{pq}\delta_{\alpha \beta} e^{-\frac{k_p}{\gamma_p} t}+\delta_{pq}\delta_{\alpha \beta} \frac{\nu f_A^2 N}{4k_p^3 }e^{-\frac{k_p (\Delta\tau+t)}{\gamma_p}}\\
&\times  \bigg[\Theta (\Delta\tau-t) \left(-\gamma_p  e^{\frac{2 \Delta\tau k_p}{\gamma_p }}+2 (\Delta\tau-t) k_p e^{\frac{k_p (\Delta\tau+t)}{\gamma_p }}+\gamma_p  e^{\frac{2 k_p t}{\gamma_p }}\right)
+\gamma_p  \left(e^{\frac{\Delta\tau k_p}{\gamma_p }}-1\right)^2\Bigg]\\
\label{eq:active_rouse_correlation}
\end{split}
\end{equation}
\end{widetext}
\noindent Thus the correlation is no longer exponential in time. The end-to-end vector of a flexible chain is  $P_{NGA}(t)=R_N^{NGA}(t)- R_0^{NGA}(t)$. Average of its magnitude's square is \cite{supplemental}, 
\begin{equation}
\begin{split}
\left<P_{NGA}^2(t)\right>&=Nb^2\\
&+\sum_{p=\textrm{odd}}^{\infty} \frac{24\nu f_A^2 N \left(\gamma_p  \left(e^{-\frac{\Delta \tau  k_p}{\gamma_p }}-1\right)+\Delta \tau  k_p\right)}{ k_p^3}
\end{split}
\end{equation}

\noindent $Nb^2$ is the average end-to-end distance in equilibrium. The swelling is defined as $(\left<P_{NGA}^2(t)\right>-Nb^2)$. 
\noindent Similarly for Gaussian active noise,  $P_{GA}(t)=R_N^{GA}(t)-R_N^{GA}(0)$ and average of its magnitude's square  \cite{supplemental},
\begin{equation}
\begin{split}
\left<P^2_{GA}(t)\right> &=Nb^2+\sum_{p=\textrm{odd}}^{\infty}\frac{24NC}{k_p\gamma_p {(\frac{k_p}{\gamma_p}+\frac{1}{\tau_A})}}\\
\end{split}
\end{equation}
\noindent From Fig.(\ref{fig:swelling})(a), it can be clearly seen that swelling factor is higher in case of Gaussian active noise in comparison to non-Gaussian, for any given number of monomers ($N$) with $N_B=1$. However, this trend could be reversed if we increase the number of bacteria $(N_B)$ or the strength of the active noise ($f_A$). In addition, from Fig.(\ref{fig:swelling})(b), it can be seen that percentage of swelling (\%swelling) in case of non-Gaussian active noise, increases with the force $(f_A)$ exerted by the bacteria. 
\noindent For a flexible chain, MSD of the vector $P_{NGA}(t)$ can also be derived from the MSD of single particle as \cite{supplemental},
\begin{widetext}
\begin{equation}
\begin{split}
\sigma_{N0}^2 (t)&= \left<\left(P_{NGA}(t)-P_{NGA}(0)\right)^2\right>
=16\sum_{p=\textrm{odd}}^{\infty}3\left[\frac{2k_B T}{k_p}+\frac{N\nu f_A^2 \left(\gamma_p  \left(e^{-\frac{\Delta \tau  k_p}{\gamma_p }}-1\right)+\Delta \tau  k_p\right)}{ k_p^3}\right]
- \frac{6k_BT}{k_p}e^{-\frac{k_p t}{\gamma_p}}\\
&- \frac{3N\nu f_A^2 }{2k_p^3 }  e^{-\frac{k_p (\Delta\tau+t)}{\gamma_p }} 
\times \Bigg[\Theta (\Delta\tau-t) \left(-\gamma_p  e^{\frac{2 \Delta\tau k_p}{\gamma_p }}+2 \Delta\tau k_p e^{\frac{k_p (\Delta\tau+t)}{\gamma_p }}-2 k_p t e^{\frac{k_p (\Delta\tau+t)}{\gamma_p }}+\gamma_p  e^{\frac{2 k_p t}{\gamma_p }}\right)
+\gamma  \left(e^{\frac{\Delta\tau k_p}{\gamma_p }}-1\right)^2\Bigg]
\label{MSD_polymer}
\end{split}
\end{equation}
\end{widetext}
\noindent To consider the center of mass (COM) motion of the polymer, we  put $k_p=0$ in Eq. (\ref{eq:rouse-mode}). 

\noindent For our active Rouse chain, MSD of COM \cite{supplemental}
\begin{widetext}
\begin{equation}
\begin{split}
\sigma_{COM}^2(t)=\left< (R_c^{NGA} (t)-R_c^{NGA} (0))^2 \right>&=\frac{6 k_B T t}{N \gamma}+\frac{3 f_A^2 N_B N \left((\Delta \tau -t)^3 \Theta (\Delta \tau -t)+\Delta \tau ^2 (3t-\Delta \tau )\right)}{6 N^2\gamma ^2 (\Delta \tau +\tau )}
\label{msd_com_rouse}
\end{split}
\end{equation}
\end{widetext}
\noindent Time evolution of active MSD for end-to-end distance exhibits growth faster thermal MSD  as shown in Fig. (\ref{fig:msd}). However, active MSD for COM shows two step growth with time. The initial growth is diffusive and it coincides with that of thermal for lower activity. At larger time scale, active MSD for COM grows faster than thermal MSD, reflecting nonequilibrium signature of the bath. For higher activity, short time MSD of COM can grow faster than that of thermal. This can be seen in the limit $t\rightarrow 0$ where $\sigma_{COM}^2(t)=\frac{6k_B T t}{\gamma}+\frac{f_A^2 N_B t^2 \Delta\tau}{2N\gamma ^2 (\Delta \tau +\tau )}$. With higher values of $f_A$, $t^2$ behavior can dominate over $t$. In other words, it is possible to observe a superdiffusive behavior of the COM at short time for higher activity. This is consistent with very recent experimental observation on active enzymes, where a short time similar superdiffusive behavior emerges out  \cite{jee2018catalytic}.  In the long time limit, MSD for COM is purely diffusive, and the effective diffusivity of COM is $D_{\text{com}}^{\text{ac}}=\frac{6k_B T}{\gamma N}+\frac{\Delta \tau ^2 f_A^2 N_B}{2 N \gamma^2 (\Delta \tau +\tau )}$ \cite{supplemental}. Experimentally measured concentration of $E. Coli$ bacteria in a biological cell is typically in the range of $10^{10} \text{cells}/\text{ml}$ \cite{maggi2014generalized}. Volume of a Rouse chain is $Nb^2$ where $N=100$ and $b=0.38 \text{nm}$. Thus $N_B \sim 1$. So in the long time limit, $\frac{D_{\text{com}}^{\text{ac}}}{D_{\text{com}}^{\text{th}}}\approx 10^5$  where parameters are chosen consistently with the real values, such as $N=100, b=0.38 \text{nm}, \gamma=9.42\times 10^{-12} \text{kgs}^{-1},  T=300K, f_A=1 pN$ \cite{samanta2016chain}. For homogenous medium $\Delta\tau=0.1 s, \tau=0.01 s$ \cite{strong1998adaptation} and $N_B=1$ \cite{supplemental}.
\noindent The time-correlation function for the vector $P_{NGA}(t)$ connecting the $N^{th}$ and the $0^{th}$ monomer can be written as $\Phi(t)=\left<P_{NGA}(t).P_{NGA}(0)\right>=16\sum_{p=\textrm{odd}}^{\infty}3\left<X_p(t)X_q(0)\right>$
and reconfiguration time which can be viewed as an effective relaxation time for the chain, is defined as \cite{samanta2013end,supplemental},
\begin{widetext}
\begin{equation}
\begin{split}
\tau_{\text{rcon}}^{\text{ac}}&=\int_{0}^{\infty}dt \frac{\Phi_{N0}(t)}{\Phi_{N0}(0)}
=\frac{16\sum_{p=\textrm{odd}}^{\infty}3\left[\frac{\gamma_p  k_B T}{k_p^2}+\frac{f_A^2 N N_B \left(2 \gamma_p^2+\Delta \tau ^2 k_p^2+2 \gamma  \Delta \tau  k_p\right)}{4 k_p^4 (\Delta \tau +\tau )}\right]}{16\sum_{p=\textrm{odd}}^{\infty}3\left[\frac{k_B T}{k_p}+\frac{f_A^2 N N_B e^{-\frac{\Delta \tau  k_p}{\gamma_p }} \left(2 \gamma_p +4 \Delta \tau  k_p e^{\frac{\Delta \tau  k_p}{\gamma_p }}\right)}{4 k_p^3 (\Delta \tau +\tau )}\right]}
\label{eq:reconfi-time}
\end{split}
\end{equation}
\end{widetext}

\begin{figure}[t]
\centering
 \includegraphics[width=9.25cm,height=!]{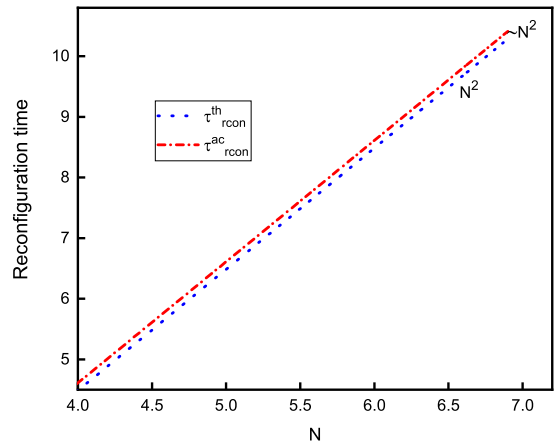}
 \caption{Log-Log plot of Reconfiguration time versus $N$ for a Rouse chain. The values of the parameters used are $b=1, \gamma=1, k_B=1, T=1, \Delta\tau=1, \tau_A=1, N_B=1, f_A=0.2$.}
 \label{fig:reconfi}
 \end{figure}

\noindent In the limit $f_A\rightarrow 0$, $\tau_{\text{rcon}}^{\text{ac}}=\tau_{\text{rcon}}^{\text{th}}=\frac{N^2 b^2 \gamma}{36 k_B T}$ which is an analytically exact result. It should be noted that reconfiguration time is related to inter-monomer distance vector while reconfiguration time in single-molecule nanosecond fluorescence correlation spectroscopy (nsFCS) is related to fluctuations of the absolute distance (distance along the backbone of a polymer). However, in Rouse model, the distance vector and absolute distance autocorrelation times are comparable  \cite{makarov2010spatiotemporal}. Recent theoretical investigations on the motion of chromosomal loci in eukaryotic nuclei reveal that Rouse model is perhaps the best possible description for bacterial chromosomes \cite{liu2015ghost}. Thus reconfiguration time can be a useful tool to measure the dynamics of activity-induced conformational changes of long chain molecules \cite{di2018anomalous}.  From Fig.  (\ref{fig:reconfi}), it is evident that the chain reconfiguration is slower in the presence of active noise. However, it scales as $~N^\sigma$, where the exponent $\sigma \approx 2$, as in case of thermal chain.  The ratio, $\frac{\tau_{\text{rcon}}^{\text{ac}}}{\tau_{\text{rcon}}^{\text{th}}}$ has practically no-dependence on parameters, such as $N_B, f_A$, $b$ or even number of monomers $(N)$ and the slowing down of the reconfiguration is in the order of $15-20\%$.

\noindent In the present work, our model of single polymer in a bath of bacteria takes care of non-Gaussian nature of random noise arising due to collisions from swimming bacteria. To the best of our knowledge, in all previous theoretical attempts to investigate the dynamics of polymer in a bath of bacteria, the active noise was modeled as Gaussian \cite{vandebroek2015dynamics,samanta2016chain,eisenstecken2017internal,osmanovic2017dynamics,ghosh2014dynamics}. Our model explicitly accounts for the run-and-tumble motion of bacteria in addition to the directionality of the kicks, using a shot noise, non-Gaussian in nature. In case, $\frac{|t-t^{\prime}|}{\tau_A}$ is sufficiently small $(<\frac{\tau_{\text{relax}}}{\tau_A})$,  a transition from Gaussian to non-Gaussian active noise is predicted which is consistent with experimental observation by Volpe $et\,\,al.$ \cite{argun2016non}. We find, MSD of COM and end-to-end distance grow faster due to collisions from active particles and can even be superdiffusive at higher activity as observed recently in case of active enzymes \cite{jee2018catalytic}. From our analysis, it can be clearly seen that the chain reconfiguration dynamics becomes slower owing to the act of activity. Surprisingly, the chain length dependence on reconfiguration dynamics remains unchanged even in the presence of active noise. We do a comparative study of swelling behavior for a single chain in the presence of Gaussian and non-Gaussian active noise especially on the value of $N_B$. However this result depends on the choice of parameters. In principle, properly designed single molecule experiments should be able to verify our results.
\\
\\
\noindent SC would like to thank Ramkishor Sharma for helpful discussions. RC acknowledges SERB for financial support (Project No. SB/SI/PC-55/2013). SC acknowledges DST-Inspire for the fellowship.



\bibliographystyle{apsrev}


\end{document}